\documentclass[twocolumn,preprintnumbers,amsmath,amssymb,prb,showpacs]{revtex4}

\usepackage{graphicx}
\usepackage{dcolumn}
\usepackage{bm}

\begin{document}

\title{Rotational motion of magnon and thermal Hall effect} 

\author{Ryo Matsumoto} 
\affiliation{Department of Physics, Tokyo Institute of Technology, Tokyo 152-8551, Japan} 
\author{Shuichi Murakami}%
\email{murakami@stat.phys.titech.ac.jp}
\affiliation{Department of Physics, Tokyo Institute of Technology, Tokyo 152-8551, Japan}

\pacs{
85.75.-d, 
66.70.-f, 
75.30.-m, 
75.47.-m 
}

\date{\today}

\begin{abstract}
Due to the Berry curvature in momentum space, the magnon wavepacket undergoes two types of orbital motions in analogy with the electron system: the self-rotation motion and a motion along the boundary of the sample (edge current). The magnon edge current causes the thermal Hall effect, and these orbital motions give corrections to the thermal transport coefficients. We also apply our theory to the magnetostatic spin wave in a thin-film ferromagnet, and derive expression for the Berry curvature. 
\end{abstract}

\maketitle

\section{Introduction}\label{sec int}
Recently in the spintronics field, the magnon (spin wave~\cite{Kittel}) transport in an insulating magnet attracts much attention as a candidate of a carrier of the spin information with good coherence and without dissipation of the Joule heating. In particular, magnon can propagate over centimeter distance in some magnets~\cite{Buttner}, e.g. yttrium-iron-garnet (YIG), and this is long enough compared with the spins in metals and doped semiconductors. The magnon current can be experimentally generated by the spin Hall effect~\cite{Kajiwara}, and its motion can be observed by the time- and space-resolved  measurement methods~\cite{Hillebrands}. Besides, a precise control of the spin information is necessary for the application in spintronics devices.

The thermal Hall effect (Righi-Leduc effect) of the magnon, which is useful to control the magnon transport, is predicted theoretically by Katsura \textit{et al.}~\cite{Katsura} and observed experimentally by Onose \textit{et al.}~\cite{Onose}. Katsura \textit{et al.} considered a ferromagnet with the Kagom\'{e} lattice structure, and calculated a thermal Hall conductivity by the Kubo formula; Onose \textit{et al.} measured the thermal Hall conductivity using an insulating ferromagnet Lu$_2$V$_2$O$_7$ which has a pyrochlore structure with the Dzyaloshinskii-Moriya (DM) interaction. On the other hand, we found in our recent study~\cite{ours} that there are correction terms to the thermal Hall conductivity in the linear response theory, and showed that the thermal Hall effect of the magnon arise from the edge current of the magnon in the semiclassical picture. Our theory is applicable not only to the quantum mechanical spin wave, e.g. in Lu$_2$V$_2$O$_7$, but also to the classical magnetostatic spin wave, where the wave length is long and exchange coupling is negligible, e.g. in YIG film.

In the present paper, we develop the transport theory of magnons with detailed calculations. Some parts of the theory has been published in Ref.~\onlinecite{ours}. There are two approaches; the semiclassical theory and the linear response theory. From the semiclassical equation of motion, the magnon edge current are described by the Berry curvature in momentum space and does not depend on the details of the system such as the shape of the boundary of the sample. From this magnon edge current, we obtain the magnon current and energy current density under a spatial variation of the temperature or the chemical potential, resulting in the thermal Hall effect of the magnon. 
Since our result of the thermal Hall conductivity does not agree with the previous works in Refs.~\onlinecite{Katsura,Onose}, we reformulated the linear response theory in analogy with the electron system\cite{ours}, by noting that the temperature gradient is not a dynamical force but a statistical force. It is identified that the difference from the previous work arises from orbital motions of the magnon, and that the magnon rotates around itself besides the magnon edge current. 

We apply our theory to Lu$_2$V$_2$O$_7$, and calculate the orbital angular momenta of the rotational motions of the magnon. 
For another application, the expression of the Berry curvature for the magnetostatic forward volume wave in YIG is derived. In this case the Berry curvature of the highest energy band enhances and that of the other bands converges to $0$ at $k=0$. Besides the Berry curvature 
becomes larger as the magnetic field becomes small.   

This paper is organized as follows. We present the semiclassical theory for the magnon and consider the thermal Hall effect of the magnon in Section~\ref{sec semi}. The linear response theory with a temperature gradient and the orbital motions of the magnon are discussed in Section~\ref{sec line}. Section~\ref{sec app} and Section~\ref{sec msw} are devoted to applications of our theory to Lu$_2$V$_2$O$_7$ and YIG, respectively. We conclude with a summary in Section~\ref{sec conc}, and a brief review of the linear response theory for the electron system and some useful equations are presented in Appendix~\ref{sec appendix}.  

Throughout this paper we consider two-dimensional insulating magnetic systems for simplicity, and assume that magnons do not interact with each other. Generalization to three-dimensional magnets is straightforward.

\section{Semiclassical theory}\label{sec semi}
Our approach is based on the semiclassical theory, in analogy with the electron system~\cite{Sundaram Niu, Xiao Niu}. We consider a magnon wavepacket which is well localized around the center $\left(\bm{r}_\text{c}, \bm{k}_\text{c} \right)$ in the phase space: 
\begin{equation}
	\left| W_n \right\rangle = \int d\bm{k}a_n(\bm{k},t)\left| \phi_{n\bm{k}}\right\rangle, 
\end{equation}
where $\left|\phi_{n\bm{k}}\right\rangle$ is the Bloch wave function in the $n$th magnon band, $a_n(\bm{k},t)$ satisfies
\begin{align}
	\int d\bm{k}\left| a_n(\bm{k},t) \right|^2 &=1, 
	\\
	\int d\bm{k} \left| a_n(\bm{k},t) \right|^2 \bm{k} &=\bm{k}_\text{c}. 
\end{align}
and $\left| W_n \right\rangle$ satisfies 
\begin{equation}
	\left\langle W_n \right| \hat{\bm{r}} \left| W_n \right\rangle = \bm{r}_\text{c}. 
	\label{eq pake}
\end{equation}
Hereafter we omit the index $c$ for brevity. 
The dynamics of the wavepacket is described by the semiclassical equation of motion, which includes the topological Berry phase term: 
\begin{align}
	\dot{\bm{r}}	&=	\frac{1}{\hbar}\frac{\partial \varepsilon_{n\bm{k}} }{\partial \bm{k}}-\dot{\bm{k}}\times\bm{\Omega}_n(\bm{k}), \label{eq eomr}\\
	\hbar \dot{\bm{k}}&=	-\nabla U(\bm{r}) \label{eq eomk}.  
\end{align}
Here $n$ is the band index, $\varepsilon_{n\bm{k}}$ is the energy of the magnon in the $n$th band, $\bm{\Omega}_n(\bm{k})$ is the Berry curvature in momentum space: 
\begin{equation}
\bm{\Omega}_n(\bm{k})=i\left\langle \frac{\partial u_n}{\partial \bm{k}}\right|\times\left| \frac{\partial u_n}{\partial \bm{k}}\right\rangle, 
\end{equation}
with $ | u_n(\bm{k}) \rangle$ being the periodic part of Bloch waves in the $n$th band defined as $\phi_{n\bm{k}}(\bm{r})=u_n(\bm{k},\bm{r})e^{i\bm{k}\cdot\bm{r}}$. $U(\bm{r})$ is a confining potential which exists only near the boundary of the sample. This potential $U(\bm{r})$ forbids the magnon wavepacket going outside of the sample, and its gradient exerts a force on magnons. Such approach of the confining potential is successful in describing the edge picture of the quantum Hall effect in electron systems~\cite{Buttiker}. Thus we have similarly introduced the confining potential for magnons. 
Strictly speaking, for the validity of Eqs.~\eqref{eq eomr} and \eqref{eq eomk}, the spatial variation of $U(\bm{r})$ should be much slower, compared with the size of the wavepacket. Nevertheless, as we can see from the quantum Hall effect as an example, many of the results for the slowly varying $U(\bm{r})$ are expected to carry over to the case of rapidly changing $U(\bm{r})$ as well. 

Near the edge of the sample, there exists an edge current of magnons due to the anomalous velocity term $-\dot{\bm{k}}\times\bm{\Omega}_n(\bm{k})=\nabla U(\bm{r}) /\hbar \times\bm{\Omega}_n(\bm{k})$ in Eq.~\eqref{eq eomr}. For example, the magnon edge current for the edge along the $y$ direction is expressed as 
\begin{align}
	I_y &=\int_a^b dx \frac{1}{V}\sum_{n,\bm{k}} \rho(\varepsilon_{n\bm{k}} +U(\bm{r})) \left[ \nabla U(\bm{r}) /\hbar \times\bm{\Omega}_n(\bm{k}) \right]_y , 
\notag \\
	&=-\frac{1}{\hbar V} \sum_{n,\bm{k}} \int_{\varepsilon_{n\bm{k}}}^{\infty} d\varepsilon 
\rho(\varepsilon) \Omega_{n,z}(\bm{k}),  \label{eq edge1}
\end{align}
where $x=a$ and $x=b$ are chosen well inside and outside of the sample so that $U(a)=0$ and $U(b)=\infty$, $V$ is the area of the sample, $\rho(\varepsilon )$ is the Bose distribution function $\rho(\varepsilon )=(e^{\beta(\varepsilon  -\mu)}-1)^{-1}$, $\beta=1/{k_{\text{B}}T}$,  $k_{\text{B}}$ is the Boltzmann constant, $\mu$ is the chemical potential, and $T$ is the temperature. 
Henceforth the magnon current means the current of the magnon number. 
We used in Eq.~\eqref{eq edge1} the fact that $\bm{\Omega}_n(\bm{k})$ in the two-dimensional system is perpendicular to the plane. Similarly we obtain the edge current for the edge along the $x$ direction $I_x$, which is identical to $I_y$. Thus the edge current does not depend on the edge direction or the expression for the confining potential $U(\bm{r})$. Therefore, the magnon moves even along the curved edge. 
Here we should note that in addition to the velocity along the edge (i.e. the second term in the r.h.s.\ of Eq.(\ref{eq eomr})), 
there exists a group velocity (the first term in the r.h.s.\ of Eq.(\ref{eq eomr})). Because of this group velocity, 
a single wavepacket does not go purely along the edge. What we have shown is that there is an additional velocity 
along the edge due to Berry curvature, and the total magnon edge current is given by Eq.~(\ref{eq edge1}) when all the magnons in thermal equilibrium 
are summed over.

\begin{figure}
\includegraphics[scale=0.4]{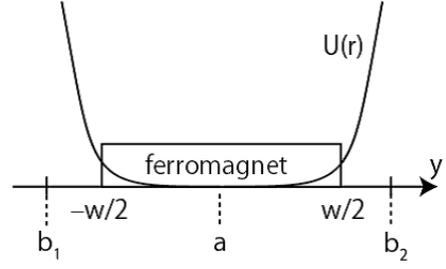}
\caption{Coordinate of the ferromagnet, used for the calculation of the edge current. $U(\bm{r})$ is a confining potential. }
\label{fig co}
\end{figure}

If the chemical potential $\mu$ or temperature $T$ varies spatially, the thermal Hall effect will occur because the magnon edge current no longer cancels  between one edge and the opposite edge, and a net current will appear. In the following we show the details and calculate thermal transport coefficients. We focus on the edge current in the $x$ direction with small temperature gradient in the $y$ direction as an example, and set the coordinate system shown in Fig.~\ref{fig co}.  
Here $w$ is the width of the system and $a$, $b_1$, $b_2$ is defined as $U(a)=0$, $U(b_1)=U(b_2)=\infty$ and $b_1<-w/2<a<w/2<b_2$. 
The current density is obtained by summing up the local current density ${j}_x(y)$ and dividing it by the width: 
\begin{align}
	j_x		=\frac{1}{w}\int_{b_1}^{b_2}dy j_x(y) = \frac{1}{w}\int_{a}^{b_2}dy j_x(y) + \frac{1}{w}\int_{b_1}^{a}dy j_x(y). 
	\label{eq jx}
\end{align}
Here we defined $j_x(y)$ as the following: 
\begin{equation}
j_x(y)=\frac{1}{\hbar V} \sum_{n,\bm{k}} \rho(\varepsilon_{n\bm{k}} +U(\bm{r});T(y))  \frac{\partial U(\bm{r})}{\partial y}  {\Omega}_{n,z}(\bm{k}) .
\end{equation}
This quantity is nonzero when $ {\partial U(\bm{r})}/{\partial y} \neq 0$, i.e., $y\sim \pm w/2$. At these points, 
\begin{align}
	&\rho(\varepsilon_{n\bm{k}} +U(\bm{r});T(y)) \\
	&\approx 
	\begin{cases}
		\rho(\varepsilon_{n\bm{k}} +U(\bm{r});T(\frac{w}{2})) \ \ \ \  (y \sim  \frac{w}{2}) ,
		\notag \\ 
		\rho(\varepsilon_{n\bm{k}} +U(\bm{r});T(-\frac{w}{2})) \ \ ( y\sim -\frac{w}{2} ) . 
	\end{cases}
\end{align}
Thus Eq.~\eqref{eq jx} is written as 
\begin{align}
	j_x&=\frac{1}{w} \frac{1}{\hbar V} \sum_{n,\bm{k}} \int_{\varepsilon_{n\bm{k}}}^{\infty} d\varepsilon 
	\notag \\
	& \ \ \ \ \times
\left(\rho\left(\varepsilon ; T\left(\frac{w}{2}\right)\right) - \rho\left(\varepsilon ;T\left(-\frac{w}{2}\right)\right) \right)\Omega_{n,z}(\bm{k})
	\notag \\
	&=\frac{\partial}{\partial y} \left( \frac{1}{\hbar V} 
	\sum_{n,\bm{k}} 
	\int_{\varepsilon_{n\bm{k}}}^{\infty} \rho(\varepsilon;T(y) ) {\Omega}_{n,z}(\bm{k}) d\varepsilon \right). 
	\label{eq jx2}
\end{align}
The edge current along the $y$ direction with temperature gradient in the $x$ direction is similarly written as
\begin{equation}
	j_y=-\frac{\partial}{\partial x} \left( \frac{1}{\hbar V} 
	\sum_{n,\bm{k}} 
	\int_{\varepsilon_{n\bm{k}}}^{\infty} \rho(\varepsilon;T(x) ) {\Omega}_{n,z}(\bm{k}) d\varepsilon \right). 
	\label{eq jy2}
\end{equation}
In the presence of the chemical potential gradient the edge current can be written as the same form like Eqs.~\eqref{eq jx2} and \eqref{eq jy2}.  
Therefore, if the spatial variation is well gradual, the edge current density is written as: 
\begin{equation}
	\bm{j}=\nabla \times \frac{1}{\hbar V} 
	\sum_{n,\bm{k}} 
	\int_{\varepsilon_{n\bm{k}}}^{\infty} \rho(\varepsilon )\bm{\Omega}_n(\bm{k}) d\varepsilon 
	\label{eq edge2_1}. 
\end{equation} 
In the same way, the energy current from the edge current density is written as 
\begin{equation}
	\bm{j}_{\text{E}}=\nabla \times \frac{1}{\hbar V} 
	\sum_{n,\bm{k}} 
	\int_{\varepsilon_{n\bm{k}}}^{\infty}\varepsilon \rho(\varepsilon )\bm{\Omega}_n(\bm{k})    d\varepsilon 
	\label{eq edge2_2}. 
\end{equation}
From Eqs.~\eqref{eq edge2_1} and \eqref{eq edge2_2}, we can derive various transverse transport coefficients. 
For instance, in the presence of the temperature gradient in the $y$ direction again, the edge current and energy current density in the $x$ direction are written as
\begin{align}
	\left( j\right)_x^{\nabla T} 	&= T \partial_y \left( \frac{1}{T} \right)   
	\sum_{n,\bm{k}} 
	\int_{\varepsilon_{n \bm{k} } }^{\infty}  \frac{\varepsilon - \mu}{\hbar V}  \left(\frac{d\rho}{d\varepsilon} \right) 
	\Omega_{n,z} (\bm{k})d\varepsilon ,
	\label{eq Tvary1}
	\\
	\left( j_{\text{E}}\right)_x^{\nabla T} 	&= T \partial_y \left( \frac{1}{T} \right)   
	\sum_{n,\bm{k}} 
	\int_{\varepsilon_{n \bm{k} } }^{\infty} \frac{\varepsilon (\varepsilon - \mu) }{\hbar V} \left(\frac{d\rho}{d\varepsilon} \right) 
	\Omega_{n,z} (\bm{k})d\varepsilon . 
\end{align} 
Here we note that the temperature gradient affects these currents through the Bose distribution function as $\partial_T \rho \left( \varepsilon \right)= \left( \varepsilon - \mu \right) \left( \partial \rho\left( \varepsilon \right)/\partial \varepsilon \right)T \partial_T \left( 1/T\right)$. 
Similarly we obtain these currents in the presence of the gradient of the chemical potential in the $y$ direction: 
\begin{align}
	\left( j	\right)_x ^{\nabla \mu} 	&= -\partial_y \mu   \frac{1}{\hbar V}
	\sum_{n,\bm{k}} 
	\int_{\varepsilon_{n \bm{k} } }^{\infty} \left(\frac{d\rho}{d\varepsilon} \right) \Omega_{n,z}(\bm{k}) d\varepsilon ,
	\\
	\left( j_{\text{E}}	\right)_x^{\nabla \mu}	&= -\partial_y \mu   \frac{1}{\hbar V}
	\sum_{n,\bm{k}} 
	\int_{\varepsilon_{n \bm{k} } }^{\infty}  \varepsilon  \left(\frac{d\rho}{d\varepsilon} \right)  \Omega_{n,z} (\bm{k})d\varepsilon . 
	\label{eq muvary2}
\end{align}
Now we define a heat current as $\bm{j}_\text{Q}\equiv\bm{j}_\text{E} - \mu \bm{j}$, and write down the linear response of the magnon current and heat current as 
\begin{align}
	\bm{j}	&=L_{11}\left[ -\nabla U	-\nabla \mu \right]
              + L_{12}\left[ T \nabla\left(\frac{1}{T}\right) \right]  
	\label {eq mag current}, 
	\\	
	\bm{j}_{\text{Q}}	&=L_{12}\left[ -\nabla U	-\nabla \mu  \right] 
                 + L_{22}\left[ T \nabla\left(\frac{1}{T}\right)  \right].  
	\label {eq heat current}
\end{align}
From Eqs.~\eqref{eq Tvary1}-\eqref{eq muvary2}, the transverse thermal transport coefficients $L^{xy
}_{ij}$ can be derived as  
\begin{align}
L^{xy}_{ij}=-\frac{1}{\hbar V\beta ^q} 
\sum_{n,\bm{k}} 
\Omega_{n,z}(\bm{k})c_q(\rho_n), 
\label{eq Lxy1}
\end{align}
where $i,j=1,2$, 
$c_q(\rho_n)
=\int_{\varepsilon _{n\bm{k}}}^{\infty}d\varepsilon
(\beta\left(\varepsilon-\mu )\right)^q
\left(-\frac{d\rho}{d\varepsilon } \right)  
=\int_0^{\rho_n} \left( \log \frac{1+t}{t} \right)^q dt 
$, 
$q=i+j-2$, and $\rho_n\equiv  \rho(\varepsilon_{n\bm{k}}) $. 
For example, $c_0(\rho)=\rho$, 
$c_1(\rho)=   \left( 1+\rho   \right) \log  \left( 1+\rho  \right) -\rho  \log \rho  $, and 
$c_2(\rho) =   \left( 1+\rho   \right) \left( \log   \frac{1+\rho}{\rho}  \right)^2 - \left(  \log \rho \right)^2 -2\text{Li}_2(-\rho) $, where $\text{Li}_2(z)$ is the polylogarithm function. Finally we derive the thermal Hall conductivity in a clean limit by substituting Eq.~\eqref{eq Lxy1} to $\kappa^{xy}={L_{22}^{xy}}/T$, 
\begin{equation} 
\kappa^{xy}=\frac{2k^2_{\text{B}}T} {\hbar V} \sum_{n,\bm{k}}  c_2(\rho_n) \text{Im} \left\langle \frac{\partial u_n}{\partial k_{x}}   \left|  \frac{\partial u_n}{\partial k_{y}} \right. \right\rangle .
\label{eq kappaxy}
\end{equation}
Thus the thermal Hall conductivity is expressed as the Berry curvature in momentum space, which is sensitive to the magnon band structure. 
Since the Berry curvature part is expressed as 
\begin{equation}
\text{Im} \left\langle \frac{\partial u_n}{\partial k_{x}}   \left|  \frac{\partial u_n}{\partial k_{y}} \right. \right\rangle 
=\sum_{m (\neq n)}  \text{Im} \frac{  \left\langle u_n \left| \frac{\partial H}{\partial k_x}\right| u_m \right\rangle   \left\langle u_m \left| \frac{\partial H}{\partial k_y}\right| u_n \right\rangle} {\left( \varepsilon_{n\bm{k}} - \varepsilon_{m\bm{k}} \right)^2} .
\end{equation}
Hence $\kappa^{xy}$ in Eq.~\eqref{eq kappaxy} is enhanced if there is an avoided band crossing.

\section{Linear response theory}\label{sec line}
Compared with our result in Eq.~\eqref{eq kappaxy}, the expression for the thermal Hall conductivity obtained in the previous works~\cite{Katsura, Onose}, 
\begin{equation}
	\bar{\kappa}^{xy}=\frac{2} {\hbar V T} \sum_{n,\bm{k}}  \rho_n \text{Im} 
	\left\langle \frac{\partial u_n}{\partial k_{x}}  \right|  \left(\frac{H+\varepsilon_{n\bm{k}}}{2}\right)^2 \left| \frac{\partial u_n}{\partial k_{y}} \right\rangle ,
\label{eq prekappa}
\end{equation}
is different in some points. As we see later in this section, in the linear response theory, our result (Eq.~\eqref{eq kappaxy}) consists of two parts $(S)^{xy} _{ij}+(M)^{xy} _{ij}$ (defined in Eqs.~(\ref{eq Sijm})-(\ref{eq M22m})), while the result in Refs.~\onlinecite{Katsura, Onose} (Eq.~\eqref{eq prekappa}) contains only $(S)^{xy} _{ij}$. The correction term $(M)^{xy} _{ij}$ comes from orbital motions of magnons. 
In the following, we apply the linear response theory to the magnon system in analogy with the electron system~\cite{Smrcka Streda, Oji Streda, Bergman, Xiao2, Thonhauser, Xiao1}. In Appendix~\ref{sec appendix}, we briefly review the linear response theory under the temperature gradient in the electron system, 
and derive some useful expression for the thermal transport coefficients. 

Here we shortly discuss the linear response theory with external fields~\cite{Smrcka Streda, Oji Streda, Bergman, Xiao2, Thonhauser, Xiao1}. In the presence of the temperature gradient, it is convenient to introduce a fictitious gravitational potential $\psi(\bm{r})$, 
which exerts a force proportional to the energy of the particle~\cite{Luttinger}. 
This is because in order to obtain the thermal transport coefficients by the linear response theory, we need to take the temperature gradient into the Hamiltonian as an external field. However, it is not possible to directly incorporate the temperature gradient into the linear response theory, since the temperature gradient is not a dynamical force which exerts force to the particles, but a statistical force which affects the particles through the distribution function. Therefore, to avoid this difficulty, the fictitious potential $\psi$, giving a dynamical force, has been introduced. As we see in Appendix~\ref{sec appendix}, the thermal transport coefficients from the temperature gradient are derived by calculating the coefficient from the gradient of the fictitious potential $\psi$. This is analogous to the situation that the transport coefficients from the gradient of the chemical potential can be obtained by calculating the coefficients from the electric field.

Now we consider the magnon system. Since the magnon has no charge, we cannot use the electric field $\bm{E}$ as an external field. Instead, we again use the gradient of the confining potential $-\nabla U(r)$, which appeared in Eq.~\eqref{eq eomk}. 
The perturbation Hamiltonian is written as $H^\prime = \sum_j U(\bm{r}_j)+\frac{1}{2}\left\{  H , \frac{1}{c^2} \sum_j \bm{r}_j \cdot \nabla  \psi (\bm{r}) \right\} $, where $\rm{r}_j$ is the position of the $j$th magnon, $H$ is the unperturbed Hamiltonian, and $\{\hat{A},\hat{B}\}=\hat{A}\hat{B}+\hat{B}\hat{A}$ represents the anticomutator. In equilibrium, the magnon current density and energy current density are written as 
\begin{align}
\bm{j}^{(0)}(\bm{r}) &=\frac{1}{2}\sum_{j}\left\{  \bm{v}_j, \delta(\bm{r}-\bm{r}_j)  \right\}, 
\\
\bm{j}_{E}^{(0)}(\bm{r})  & =\frac{1}{2}\left\{  H , \bm{j}^{(0)}(\bm{r})  \right\}, 
\end{align}
where $\bm{v}_j$ is the velocity operator of the $j$th magnon. In the presence of the external fields $H^\prime$, they acquire additional terms, 
\begin{align}
		\bm{j}(\bm{r})	&=\bm{j}^{(0)}(\bm{r})+\frac{1}{2}\left\{ \bm{j}^{(0)}(\bm{r}) , \frac{1}{c^2}\sum_j \bm{r}_j \cdot \nabla  \psi (\bm{r})\right\}, \\
	\bm{j}_E(\bm{r})	&=\bm{j}^{(0)}_{E}(\bm{r})+\frac{1}{2}\sum_j\left\{ U (\bm{r}_j) , \bm{j}^{(0)}(\bm{r}) \right\}  \notag \\
							&+\frac{1}{4c^2} \sum_j \left(    \left\{   \left\{ H,\bm{r}_j \cdot \nabla \psi(\bm{r}) \right\}  ,  \bm{j}^{(0)}(\bm{r})  
								\right\} \right.
							\notag \\
							&+ \left.  \left\{   \left\{ \bm{j}^{(0)}(\bm{r}),\bm{r}_j \cdot \nabla \psi(\bm{r}) \right\}  ,  H   \right\} \right)          ,
\end{align} 
where $c$ is the speed of light. 
Correspondingly, 
the thermal transport coefficients consist of two parts: $(L)^{\alpha\beta} _{ij}=(S)^{\alpha\beta} _{ij}+(M)^{\alpha\beta} _{ij}$. Here $\alpha, \beta=x,y$, $i, j=1,2$, and in this section the thermal transport coefficients $(L)^{\alpha\beta} _{ij}$ are defined as: 
\begin{align}
		\bm{J}&=(L\text)_{11}\left[ -\nabla U-{T}\nabla \left(\frac{\mu}{T} \right) \right]
              + (L\text)_{12}\left[ T \nabla\left(\frac{1}{T}\right) - \frac{\nabla\psi}{c^2} \right]  
	\label {eq current}, 
	\\	
	\bm{J}_E&=(L\text)_{12}\left[ -\nabla U-{T}\nabla \left( \frac{\mu}{T} \right) \right] 
                 + (L\text)_{22}\left[ T \nabla\left(\frac{1}{T}\right) - \frac{\nabla \psi}{c^2} \right] . 
\end{align}
A deviation of the distribution function from the equilibrium state generates $(S)^{\alpha\beta} _{ij}$, which is calculated by the Kubo formula; a deviation of the current operator due to external fields from the equilibrium state generates $(M)^{\alpha\beta} _{ij}$.  
In the clean limit they are expressed as 
\begin{widetext}
\begin{align}
	(S^B)^{\alpha \beta}_{ij}  	&=	\frac{i\hbar}{V} \int \rho(\eta) \text{Tr} \left( j^{\alpha}_i \frac{dG^+}{d\eta} j^{\beta}_j  \delta (\eta-H) -  j^{\alpha}_i  \delta (\eta-H)  j^{\beta}_j  \frac{dG^-}{d\eta} \right) d\eta , \label{eq Sijm} \\
	(M^B)^{\alpha \beta}_{11}  	&=	0 , 
	\ \ \ 
	(M^B)^{\alpha \beta}_{12}  	=	\frac{1}{2V} \int \rho(\eta) \text{Tr} [\delta (\eta-H) (r^{\alpha}v^{\beta}-r^{\beta}v^{\alpha})] d\eta 
	\label{eq M12m}, \\
	(M^B)^{\alpha \beta}_{22}  	&=	 \frac{1}{V}\int \eta \rho(\eta) \text{Tr} \delta (\eta-H) (r^{\alpha}v^{\beta}-r^{\beta}v^{\alpha}) d\eta 
										+ \frac{i\hbar}{4V}\int \rho(\eta) \text{Tr} \delta (\eta-H) [v^{\alpha} , v^{\beta}] d\eta \label{eq M22m}. 
\end{align}
\end{widetext}
Here $G^{\pm}$ is the Green's function $G^{\pm}(\eta)=(\eta-H\pm i\epsilon )^{-1}$ with $\epsilon$ being the positive 
infinitesimal, $\rho(\eta)$ is the Bose distribution function $\rho(\eta)=\left(e^{\beta\left(\eta-\mu \right)}-1\right)^{-1}$, $\bm{j}_{1}=\bm{v}$, $\bm{j}_{2}=\frac{1}{2}(H\bm{v}+\bm{v}H)$, $\bm{v}$ is the velocity of magnons, and the label of the superscript ``B" means a boson. 
By using Eqs.~\eqref{eq SA12}-\eqref{eq Trrv}, these thermal transport coefficients for the magnon system can be written by the wave function of magnons:
\begin{align}
	(S^B)^{\alpha\beta} _{ij} 		&= 	\frac{2}{\hbar V} \text{Im} \sum_{n,\bm{k}}  \rho_n  \left\langle \frac{\partial u_n}{\partial k_{\alpha}} \left|   \left( \frac{H+\varepsilon _{n\bm{k}}}{2}\right)^q    \right|  \frac{\partial u_n}{\partial k_{\beta}} \right\rangle , 
	\label{eq Sij 2} 
	\\
	(M^B)^{\alpha\beta} _{ij}  	&=	-	(S^B)^{\alpha\beta} _{ij} 
	\notag \\
		&+	\frac{2\left(   k_{\text{B}}T   \right)^q}{\hbar V}  \text{Im} \sum_{n,\bm{k}}  
   c_q(\rho_n)  
\left\langle \frac{\partial u_n}{\partial k_{\alpha}}   \left|  \frac{\partial u_n}{\partial k_{\beta}} \right. \right\rangle 	, 
	\label{eq Mij 2}
\end{align}
where  we have taken $\mu=0$ since the magnon number is not conserved. We note that $q=i+j-2$ and $(M^B)^{\alpha\beta} _{11}=0 $. Thus the total thermal transport coefficients are written as 
\begin{align}
	&(L^B)^{\alpha\beta} _{ij}
	=(S^B)^{\alpha\beta} _{ij}+(M^B)^{\alpha\beta} _{ij} , 
	\notag \\
	&=-\frac{\left( k_{\text{B}}T \right)^q}{\hbar V}  \sum_{n,\bm{k}} c_q(\rho_n) \Omega_{n,z}({\bm{k}}) , 
\end{align}
and from this equation we again derive the same thermal Hall conductivity as Eq.~\eqref{eq kappaxy}. 
The result in Refs.~\onlinecite{Katsura, Onose}, shown in Eq.~\eqref{eq prekappa}, contains only the contribution from $(S^B)^{\alpha\beta} _{22}$. 
Therefore the difference between the results of Refs.~\onlinecite{Katsura, Onose} and ours arise from the correction terms $(M^B)^{\alpha\beta} _{ij} $.

As we can see from Eqs.~\eqref{eq M12m} and \eqref{eq M22m}, these correction terms $(M^B)^{\alpha\beta} _{ij} $ are related to the orbital motion of the particle, namely, a reduced orbital angular momentum $\langle \bm{r} \times \bm{v} \rangle$. Equation.~\eqref{eq Mij 2} means that $(M^B)^{\alpha\beta} _{ij} $ are expressed as the Berry curvature in momentum space, which is generally nonzero. Hence, in this case, the magnon has finite orbital angular momentum due to the Berry curvature. 
This orbital angular momentum consists of two parts: the edge current and the self-rotation motion of the wavepacket. The reduced angular momentum for the edge current per unit area is derived from Eq.~\eqref{eq edge1},  
\begin{equation}
l^{\text{edge}}_z=-\frac{2}{\hbar V}
\sum_{n,\bm{k}} 
\int_{\varepsilon_{n\bm{k}}}^{\infty}d\varepsilon 
\rho(\varepsilon)\Omega_{n,z}(\bm{k}), \label{eq L edge}
\end{equation}
and that for the self-rotation motion is calculated in analogy with the electron system \cite{M C Chang} as
\begin{equation}
	l_z^{\text{self}}	=	-\frac{2}{\hbar V} \text{Im} \sum_{n,\bm{k}}  \rho_n \left\langle \left. \frac{\partial u_n}{\partial k_x} \right|  \left( H-\varepsilon_{n\bm{k}} \right)   \left|  \frac{\partial u_n}{\partial k_y} \right. \right\rangle  
\label{eq L self}.
\end{equation} 
It is easy to show that $l^{\text{edge}}_z+l_z^{\text{self}}= 2(M^B)^{xy} _{12} $. This result is expected from the Eq.~\eqref{eq M12m}, i.e., the correction terms comes from the orbital angular momentum of the magnon.




Therefore, due to the Berry curvature, the magnon generally has a nonvanishing orbital angular momentum in equilibrium. 
This orbital motion of magnon can be regarded as a generalized cyclotron motion.  
However, since the magnon has no charge, it feels no Lorentz force and cannot have a cyclotron motion
in the same sense as that of electrons. In this respect, this motion purely reflects the magnon band structure. A similar effect can be found in various 
wave phenomena such as electrons\cite{Xiao Niu}, 
photons \cite{Onoda}, and so on. 

As mentioned earlier, within the semiclassical theory, the result for the edge current is derived under the assumption that 
the spatial variation of the confining potential is slow. Nevertheless, as we have seen in this section, 
the linear response theory, which does not need 
assumptions for confining potential, gives the same transport coefficients as the semiclassical theory. 
This strongly suggests that the edge-current picture carries over to the abrupt spatial variation of the confining potential. 
In the quantum Hall systems, this idea is indeed true; at the abrupt edges of the quantum Hall system, the electrons 
undergo a skipping motion. Namely, near the edge, electrons undergo a cyclotron motion and when electrons hit the edge
they are bounced. As a whole the electrons go along the edge with skipping orbitals, which are regarded as the 
chiral edge current in the quantum Hall system. 
Therefore, we can similarly expect that in the ferromagnets with edges, the magnon will undergo a superposition of a skipping 
motion along the edge and a motion along the group velocity of the magnon.

The coherence length of the magnons is important in the orbital motions and transport of magnons. 
For the validity of the linear response theory developed above, it is implicitly assumed that the 
coherence length of the magnons is sufficiently short compared with the system size. 
By this assumption, when we apply temperature difference between the two opposite sides of the system, 
we can define a local temperature, 
and the temperature gradient becomes uniform. The linear response theory is then justified. 
Otherwise, when the coherence length is as long as the system size, the magnon transport is 
described in the similar way as the Landauer formula, and the linear response theory no longer applies.




Orbital motions of electrons generate magnetic moments by their charge. 
On the other hand, magnons do not have charge, but have magnetic moments. 
Therefore we can regard the rotating magnon as a circulating spin current.  
As is similar to the  magnetoelectric effect in noncollinear spin structure \cite{KatsuraME}, the rotation of the magnon is expected to generate an electric polarization. For this effect, the spin-orbit coupling interaction, such as Dzyaloshinskii-Moriya (DM) interaction, is necessary.

\section{Example 1: Lu$_2$V$_2$O$_7$}\label{sec app}
\begin{figure}
\includegraphics[scale=0.5]{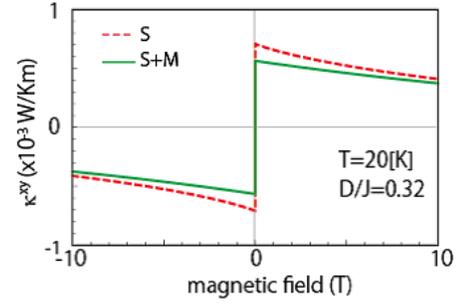}
\caption{(Color online) Dependence of the thermal Hall conductivity on a magnetic field. The red (broken) curve denotes the result which is calculated from only $(S)^{\alpha\beta} _{ij}$ calculated in Ref.~\onlinecite{Onose}; the green (solid) curve denotes the result which is calculated from $(S)^{\alpha\beta} _{ij}+(M)^{\alpha\beta} _{ij}$. } 
\label{fig compare}
\end{figure}

In this section we apply our results to the ferromagnetic Mott-insulator $\text{Lu}_2\text{V}_2\text{O}_7$ with pyrochlore structure, for which the thermal Hall effect has been measured and analyzed in Ref.~\onlinecite{Onose}. Following Ref.~\onlinecite{Onose}, 
we briefly review the magnetic properties of the material. The magnetization comes from spin-1/2 $\text{V}^{4+}$ ions with the DM interaction. The ground state is a collinear ferromagnet, because at the ferromagnetic ground state the total DM vectors for the six 
bonds sharing a single site is zero~\cite{Onose}. The DM interaction affects the spin-wave dispersion, and the 
effective spin-wave Hamiltonian is written as 
$H_{\text{eff}}=\sum_{\langle i,j \rangle}-J\bm{S}_i \cdot \bm{S}_j + \bm{D}_{ij} \cdot \left( \bm{S}_i \times \bm{S}_j \right) -g\mu_{\text{B}}\bm{H}\cdot \sum_i\bm{S}_i$,
where $\langle i,j \rangle$ denotes the nearest neighbor pairs, $J$ is the exchange interaction, $\bm{D}$ is the DM vector, $g$ is the g-factor, $\mu_{\text{B}}$ is Bohr magneton, and $\bm{H}$ is the magnetic field in the $z$ direction. 
The temperature is assumed to be much lower than the Curie temperature $T_C=70[\text{K}]$, 
for existence of well-defined Bloch waves of magnons. 
There are four magnon bands, and the lowest band is well separated from the other higher bands, with the separation much 
larger than $k_B T$. 
Actually, the differences of the energy between the lowest band and other bands near $k=0$ are written as $\varepsilon_{2}-\varepsilon_{1}\simeq 4JS\sqrt{1+f(\bm{k})}\simeq 8JS$ and $\varepsilon_3-\varepsilon_{1}=\varepsilon_{4}-\varepsilon_{1}\simeq 4JS+2JS\sqrt{1+f(\bm{k})}\simeq 8JS$, where $f(\bm{k})=\cos(2k_xA)\cos(2k_yA)+\cos(2k_yA)\cos(2k_zA)+\cos(2k_zA)\cos(2k_xA)$ and $8JS\simeq 13.6[\text{meV}]$. 
Therefore, the contribution from the lowest band is dominant, whose Berry curvature is 
$\Omega_{1,z}\simeq -\frac{A^4}{8\sqrt{2}}\frac{D}{J}\frac{H_z}{H}(k_x^2+k_y^2+2k_z^2)$ as calculated in Ref.~\onlinecite{Onose},
with $A$ being a quarter of the lattice constant. 
We can estimate the orbital angular momentum of the magnon from both the self-rotation motion $L_z^{\text{self}}$  and the edge current $L_z^{\text{edge}}$. Near $k=0$, the lowest-band dispersion is quadratic 
and we can introduce the effective mass of the magnon of the lowest band $m_1^*$, defined as $m^*_n\equiv \hbar^2(\partial^2 \varepsilon_{n\bm{k}}/\partial k^2)^{-1}$. 
The orbital angular momentum of the self-rotation motion is analytically calculated from Eq.~\eqref{eq L self}: 
\begin{align}
	 &L_z^{\text{self}} \simeq m_1^* l_z^{\text{self}} = -\frac{16 JS m^*_1}{\hbar V}\text{Im}\sum_{\bm{k}} 
\rho(\varepsilon_{1\bm{k}})\left\langle \frac{\partial u_1}{\partial k_{\alpha}} \left|  \frac{\partial u_1}{\partial k_{\beta}} \right. \right\rangle 
\notag \\
&=-\frac{ JSm^*_1}{\hbar A}\frac{D}{J}\frac{1}{24\pi^2}\left(\frac{k_{\text{B}}T}{JS}\right)^{5/2}
\int_0^\infty \frac{x^{3/2}}{e^{\left(x+\beta g \mu_{\text{B}}H\right)}-1}dx
\notag \\
&=-\frac{JSm_1^*}{\hbar A}  \frac{D}{J} \frac{1}{32\pi^{3/2}}\left(\frac{k_\text{B}T}{JS}\right)^{5/2} \text{Li}_{\frac{5}{2}}\left(e^{-\frac{ g \mu_{\text{B}} H}{k_\text{B}T}}\right). \label{eq Lz self LuVO} 
\end{align}
We obtain $L_z^{\text{self}} \simeq -0.009 \hbar$ and $L_z^{\text{edge}}=m_1^*l^{\text{edge}}_z \simeq +0.008\hbar $ per unit cell. 
The thermal Hall conductivity $\kappa^{xy}$ is also calculated, by assuming that the contribution of the lowest band dominates. 
Figure \ref{fig compare} show the result of the thermal Hall conductivity which is calculated from $(S)^{\alpha\beta} _{ij}+(M)^{\alpha\beta} _{ij}$ (solid curve) and $(S)^{\alpha\beta} _{ij}$ (broken curve).  
They correspond to our results and the previous results in Ref.~\onlinecite{Onose}, respectively.  
Our result (solid curve in Fig.~\ref{fig compare}) roughly agrees with the experimental data 
in Ref.~\onlinecite{Onose}. 

\section{Example 2: Magnetostatic spin wave}\label{sec msw}
In the following, we apply our theory to the magnetostatic spin waves in a ferromagnet. In the magnetostatic spin wave, 
the wavelength is sufficiently long and the exchange coupling between spins is negligible. 
The magnetic anisotropy comes from the demagnetizing field determined by the sample shape. 
This magnetic anisotropy due to the demagnetizing field plays the similar role as the spin-orbit coupling in electronic systems.  
It then induces the Berry curvature, and the Hall effect of spin waves appears. 

\begin{figure}
\includegraphics[scale=0.4]{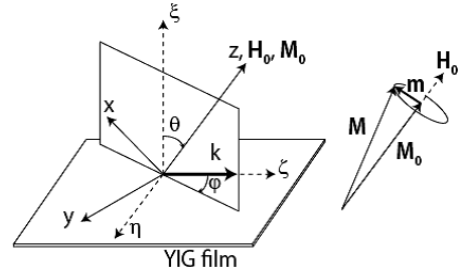}
\caption{Geometry of the coordinate axes. The magnetization $\bm{M}$ precesses around $\bm{M}_0$.  }
\label{fig MSSW}
\end{figure}

Let us consider the yttrium-iron-garnet (YIG) film which is magnetized by an external magnetic field. 
We introduce two coordinate systems $xyz$ and $\xi \eta \zeta $, shown in Fig.~\ref{fig MSSW}. The film is taken to be infinite in the $\eta$- and $\zeta$-direction, and perpendicular to the $\xi$ direction. $\zeta$ axis is chosen to be along the magnon wave vector $\bm{k}$. The $z$ direction is parallel to the saturation magnetization $\bm{M}_0$ and the internal static magnetic field $\bm{H}_0$. 
We assume that the spin wave mode has a form of the plane wave: $\bm{m}(\xi,\zeta ,t)=\bm{m}(\xi)\exp(i(k\zeta-\omega t))$, where $\omega$ is a frequency of the spin wave . The equation of motion of the magnetization is written as the following integral equation~\cite{Kalinikos}: 
\begin{equation}
 \omega_H \bm{m}(\xi) -\omega_M\int_{-L/2}^{L/2}d\xi^{\prime} \hat{G}(\xi, \xi^{\prime}) \bm{m}(\xi^{\prime})=\omega \sigma_y \bm{m}(\xi). 
 \label{eq YIG}
\end{equation}
Here we use the SI units, $L$ is the thickness of the film,  $\sigma_y=\left( \begin{smallmatrix} 0 & -i \\ i & 0\end{smallmatrix}\right)$ is the Pauli matrix, $\bm{m}(\xi)=\left( \begin{smallmatrix}m_x(\xi) \\ m_y(\xi) \end{smallmatrix} \right)$ is the vector Fourier amplitude of the spin wave which is perpendicular to $\bm{M}_0$, $\omega_H=\gamma H_0$, $\omega_M=\gamma M_0$, and $\gamma$ is the gyromagnetic ratio. 
$\hat{G}(\xi, \xi^{\prime})$ is the $2\times2$ complex matrix of the Green's function: 
\begin{align}
&\hat{G}(\xi, \xi^{\prime}) =	
	\begin{pmatrix}
		G_{xx} &G_{xy} \\ G_{yx} &G_{yy}
	\end{pmatrix},
\\
&	G_{xx}=(G_P - \delta(\xi-\xi^{\prime}))\sin^2\theta -iG_Q\sin 2\theta\cos\varphi 
\notag \\
&\ \ \ \ \ \ \ \ \ -G_P\cos^2 \theta \cos^2 \varphi ,\\
& G_{xy}=G_{yx}
\notag \\
&\ \ \ \ \ =-iG_Q\sin \theta \sin\varphi 	-G_P\cos \theta \sin \varphi \cos \varphi, \\
&G_{yy}=-G_P \sin^2\varphi ,
\end{align}
where
\begin{align}
	G_P&=\frac{k}{2}\exp (-k |\xi-\xi^{\prime}|) , \\
	G_Q&=G_P\text{sign}(\xi -\xi^{\prime}), 
\end{align}
and $\theta$, $\varphi$ are the spherical coordinates of $\bm{M}_0$ in the $\xi\eta\zeta$-space (see Fig.~\ref{fig MSSW}). 
We note that we adopt the definitions of $\theta$ and $\varphi$ used in Ref.~\onlinecite{Kalinikos}, and they are different from the standard definition of the spherical coordinates. 
The integral equation~\eqref{eq YIG} is equivalent to the linearized Landau-Lifshitz equation $d\bm{M}/dt=-\gamma (\bm{M}\times\bm{H})$, Maxwell equation in the magnetostatic limit $\nabla \times \bm{H}=0$, $\nabla \cdot \bm{B}=0$, and the usual boundary conditions for $\bm{H}$ and $\bm{B}$. Since the equation \eqref{eq YIG} is a generalized eigen value problem, we have to modify the prescription of our theory of the Berry curvature. Similar to the previous work~\cite{Chuanwei}, the Berry curvature is defined as 
\begin{equation}
\Omega_{n, \gamma}(\bm{k})=i\epsilon_{\alpha \beta\gamma}\left\langle \frac{\partial \bm{m}_{n,\bm{k}}}{\partial k_{\alpha}}  \right|   \sigma_y  \left|       \frac{\partial \bm{m}_{n,\bm{k}}}{\partial k_{\beta}} \right\rangle, 
\label{eq YIGBC}
\end{equation}
where $\epsilon_{\alpha \beta\gamma}$ is the antisymmetric tensor, $n$ is the band index of the spin wave mode, and the bra-ket product means an usual inner product of vectors and integral over $z$. 

In some cases, Eq.~\eqref{eq YIGBC} becomes zero because of the symmetry of the system. This occurs when the saturation magnetization $\bm{M}_0$ is in the film ($\theta=\pi/2$). When $\theta=\pi/2$, we can show $\Omega_{n, \gamma}(\bm{k})=0$ explicitly by performing a gauge transformation $\bm{m}^\prime \equiv U^{-1} \bm{m} = \begin{pmatrix} 1& 0 \\ 0 & i\end{pmatrix} \bm{m}$. Then Eq.~\eqref{eq YIG} becomes a generalized eigen value problem with real coefficients: 
\begin{equation}
	 \omega_H \bm{m}^\prime (\xi) - \omega_M\int_{-L/2}^{L/2}d\xi^{\prime} \hat{G}^\prime(\xi, \xi^{\prime}) \bm{m}^\prime(\xi^{\prime})=-\omega \sigma_x \bm{m}^\prime(\xi),
\label{eq gauge}
\end{equation}
where $\hat{G}^\prime(\xi, \xi^{\prime})$ is 
\begin{align}
	&\hat{G}^\prime(\xi, \xi^{\prime})=U^{-1} \hat{G}(\xi, \xi^{\prime}) U
=
\begin{pmatrix}
G^\prime_{xx} &G^\prime_{xy} \\ G^\prime_{yx} &G^\prime_{yy}
\end{pmatrix},
\\
&	G^\prime_{xx}=G_P - \delta(\xi-\xi^{\prime}) , \ \ G^\prime_{xy}=-G_Q\sin\varphi ,\\
&G^\prime_{yx}=G_Q\sin\varphi , \ \ G^\prime_{yy}=-G_P \sin^2\varphi .
\label{eq gaugetr}
\end{align}
Since all the terms in Eq.~\eqref{eq gauge} are real, the eigen vector $\bm{m}^\prime$ is also real. Correspondingly, the Berry curvature Eq.~\eqref{eq YIGBC} becomes
\begin{align}
\Omega_{n, \gamma}(\bm{k}) &=i\epsilon_{\alpha \beta\gamma}\left\langle \frac{\partial \bm{m}^\prime_{n,\bm{k}}}{\partial k_{\alpha}}  \right|  U^{-1} \sigma_y   U   \left|       \frac{\partial \bm{m}^\prime_{n,\bm{k}}}{\partial k_{\beta}} \right\rangle,  \\
&=\epsilon_{\alpha \beta\gamma}\text{Im}\left\langle \frac{\partial \bm{m}^\prime_{n,\bm{k}}}{\partial k_{\alpha}}  \right|   \sigma_x      \left|       \frac{\partial \bm{m}^\prime_{n,\bm{k}}}{\partial k_{\beta}} \right\rangle.   
\end{align}
Because $\bm{m}^\prime$ is real and there is no imaginary part, this Berry curvature vanishes. Thus when $\bm{M}_0$ is in the film, we cannot expect either an orbital rotational motion of spin wave packet or the thermal Hall effect due to the Berry curvature effect. In other words, in the magnetostatic backward volume wave (MSBVW) and the magnetostatic surface wave (MSSW), the effects of the Berry curvature do not appear. 

\begin{figure}
\includegraphics[scale=0.35]{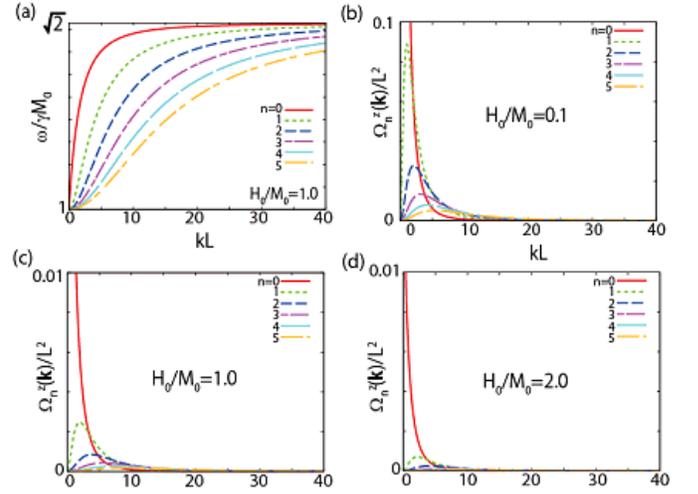}
\caption{(Color online) (a) Dispersion relation for the MSFVW mode with $n=0,1, \dots,5$, and the Berry curvature for the MSFVW mode for (b) $H_0/M_0=0.1$, (c) $H_0/M_0=1.0$, and (d) $H_0/M_0=2.0$. }
\label{fig BCYIG}
\end{figure}

On the other hand, the Berry curvature is finite if the saturation magnetization is perpendicular to the film ($\theta=0$), i.e., in the magnetostatic forward volume wave (MSFVW).  In the following, we demonstrate the calculation of the Berry curvature for MSFVW. We note that $\xi$ coincides with $z$ direction when $\theta=0$. The solution of the integral equation~\eqref{eq YIG} of the $n$-th band  for $\theta=0$ is written as~\cite{Damon}
\begin{align}
	\bm{m}_{n\bm{k}}(z)&=
	\begin{pmatrix}
		m_{n\bm{k}}^{x}(z) \\
		m_{n\bm{k}}^{y}(z)
	\end{pmatrix}
	\notag \\
	&=
	{\sqrt{N}}
	\begin{pmatrix}
		i\kappa & \nu \\
		-\nu & i\kappa
	\end{pmatrix}
	\begin{pmatrix}
		k_x \\ k_y
	\end{pmatrix}
	\cos \left( \sqrt{p} kz + \frac{n\pi}{2} \right), 
	\label{eq solfw}
\end{align}
where $\kappa=\omega_M \omega_H / (\omega^2_H-\omega^2_n)$, $\nu=\omega_M \omega_n / (\omega^2_H-\omega^2_n)$, $p=-1-\kappa>0$, $\omega_n$ is the $n$-th band energy for $n=0, 1, 2, \dots$, which is determined by 
\begin{equation}
\sqrt{p}\tan\left(\frac{\sqrt{p}kL+n\pi}{2} \right)=1, 
\label{eq eigenfreq}
\end{equation}
and $N$ is a normalization factor which is determined by 
\begin{equation}
\left\langle \bm{m}_{n,\bm{k}}  \right|   \sigma_y  \left|   \bm{m}_{n,\bm{k}} 
\right\rangle=1.
\label{eq normal}
\end{equation}
To obtain Eq.~\eqref{eq solfw}, we have rewritten the solution in Ref.~\onlinecite{Damon} in the polar coordinate into the form of the plane wave. 
The dispersion determined by (\ref{eq eigenfreq}) is shown in Fig.~\ref{fig BCYIG}(a) for $H_0/M_0=1.0$.
We use the normalization Eq.~(\ref{eq normal}), because $\bm{m}_{n,\bm{k}}^{\dagger}\sigma_y \bm{m}_{n,\bm{k}} $ is proportional 
to the 
energy density for the magnon~\cite{Fishman}.

Substituting this solution to Eq.~\eqref{eq YIGBC}, we can can calculate the Berry curvature for the $n$-th MSFVW mode. For simplicity of the notation, we set $F_n(k,z)\equiv {\sqrt{N}} \cos \left( \sqrt{p} kz + \frac{n\pi}{2} \right)$. Then the Berry curvature is written from Eq.~\eqref{eq YIGBC}, 
\begin{align}
	&\Omega_{n, z}(\bm{k})/2 \notag \\
	&=\text{Re} \left[ 
	\left\langle \frac{\partial m^x_{n,\bm{k}}}{\partial k_{x}}\right.\left|       \frac{\partial m^y_{n,\bm{k}}}{\partial k_{y}} \right\rangle
	-\left\langle \frac{\partial m^x_{n,\bm{k}}}{\partial k_{y}}\right.\left|       \frac{\partial m^y_{n,\bm{k}}}{\partial k_{x}} \right\rangle
	\right]
	\notag \\
	&=		(\kappa^2+\nu^2) k \int_{-L/2}^{L/2} dz
\frac{\partial F_n}{\partial k} F_n
	\notag \\
			&
			\ \ \ +\left\{
			k \left(
				\kappa \frac{\partial \kappa}{\partial k} + \nu \frac{\partial \nu}{\partial k} 
			\right)
			+\kappa^2+\nu^2
		\right\} \int_{-L/2}^{L/2} dz
F_n^2	.		\label{eq integ1}
\end{align}  
Using the normalization condition of Eq.~\eqref{eq normal}, one obtain  
\begin{equation}
{2\kappa \nu k^2 }\int_{-L/2}^{L/2} 
 dz  F_n^2 =1. 
\label{eq integ2}  
\end{equation}
Derivative of Eq.~\eqref{eq integ2} in terms of $k$ leads to 
\begin{equation}
	\int_{-L/2}^{L/2} dz
F_n \frac{\partial F_n}{\partial k} = -
	\left( 
		\frac{1}{2\kappa} \frac{\partial \kappa}{\partial k} + \frac{1}{2\nu} \frac{\partial \nu}{\partial k} 
+ \frac{1}{k}
	\right) \int_{-L/2}^{L/2} F_n^2 dz. 
\end{equation}
Thus Eq.~\eqref{eq integ1} is rewritten as
\begin{equation}
	\Omega_{n, z}(\bm{k})/2 =\frac{1}{4\kappa \nu k}\frac{\omega_M^2}{\omega_H^2 - \omega_n^2} \left( 
	\frac{1}{\kappa}\frac{\partial \kappa}{\partial k} - \frac{1}{\nu}\frac{\partial \nu}{\partial k}
	\right). 
\end{equation}
Since $\kappa$ and $\nu$ satisfy the following relation
\begin{equation}
\frac{1}{\kappa}\frac{\partial \kappa}{\partial k} - \frac{1}{\nu}\frac{\partial \nu}{\partial k}=-\frac{1}{\omega_n}\frac{\partial \omega_n}{\partial k}, 
\end{equation}
the Berry curvature is derived as 
\begin{equation}
\Omega_{n,z}(\bm{k})=\frac{1}{2\omega_H}\frac{1}{k}\frac{\partial \omega_n}{\partial k} \left( 1-
\frac{\omega_H^2}{\omega_n^2}\right). 
\label{eq YIGBC2}
\end{equation}
Figure \ref{fig BCYIG}(b)-(d) shows the numerical results of Eq.~\eqref{eq YIGBC2} for various magnitude of the magnetic field. 
It is surprising that the Berry curvature for any MSFVW mode is always positive, because $\omega_H<\omega_n$ and the group velocity $\partial \omega/\partial k$ is positive. In the vicinity of $k=0$, we can calculate asymptotic forms of the Berry curvature. When $k\sim 0$, $\omega_n$ is close to $\omega_H$. If we set $\omega_n=\omega_H+\Delta \omega_n$, $p$ can be written as $p\simeq \omega_M/2\Delta\omega_n \gg 1$ since $\Delta\omega_n$ is small near $k=0$. Using an approximation $\tan x\simeq x \ (x\ll 1)$ , we find 
\begin{align}
	\Delta\omega_n=
	\begin{cases}
		\dfrac{1}{4}\omega_M k L &(n=0)
		\\
		\dfrac{\omega_M}{2}\left( \dfrac{kL}{n\pi}\right)^2 &( n>0)
	\end{cases}
	\label{eq asymdis}
\end{align}
Therefore, the Berry curvature near $k=0$ can be obtained from Eq.~\eqref{eq YIGBC2} and \eqref{eq asymdis}: 
\begin{align}
	\Omega_{n,z}(\bm{k})/L^2\simeq 
	\begin{cases}
		 \left(\dfrac{1}{4} \dfrac{M_0}{H_0}\right)^2&(n=0)
		\\
		 \dfrac{1}{2}\left(\dfrac{1}{n\pi}\right)^4 \left(\dfrac{M_0}{H_0}\right)^2 (kL)^2&( n>0)
	\end{cases}
	\label{eq asymBC}
\end{align}
It is easy to see that $\Omega_{n,z}(\bm{k}=0)= 0$ for $n>0$ mode, and that for $n=0$ mode $\Omega_{0,z}(\bm{k})$ enhances up to $\left(\frac{L}{4}\frac{M_0}{H_0}\right)^2$ but does not diverge at $k=0$.

\section{Conclusions}\label{sec conc}
In summary, we found that magnon wavepacket has two types of orbital motions due to the Berry curvature in momentum space: the magnon edge current and the self-rotation motion. The magnon edge current causes the thermal Hall effect of magnon, and the self-rotation motion of magnon without Lorentz force is expected to accompany an electric polarization.  
We showed that our theory is applied to not only the exchange spin wave (quantum-mechanical magnon) e.g.~in 
Lu$_2$V$_2$O$_7$, but also the classical 
magnetostatic waves e.g.~in YIG.
In both cases, the Berry curvature is enhanced near the band crossings, where the magnon frequency in a focused band is close to those of other bands. We expect to control the Berry curvature by designing magnonic crystals~\cite{magnonics}.

\begin{acknowledgments}
We would like to thank B.~I.~Halperin, Q.~Niu, T.~Ono, and E.~Saitoh for discussions.
This work is partly supported  
by Grant-in-Aids  
from MEXT, Japan 
 (No.~21000004 and 22540327), 
and by the Global
Center of Excellence Program by MEXT, Japan through the
"Nanoscience and Quantum Physics" Project of the Tokyo
Institute of Technology.
\end{acknowledgments}

\appendix
\section{}
\label{sec appendix}
Here we briefly review the linear response theory for the electron system with a temperature gradient, developed in Refs.~\onlinecite{Smrcka Streda, Oji Streda, Bergman, Xiao2, Thonhauser}. 
In equilibrium, the electric current operator $\bm{j}^{(0)}(\bm{r})$ and the energy current operator $\bm{j}_{E}^{(0)}(\bm{r})$ are given by 
\begin{align}
\bm{j}^{(0)}(\bm{r}) &=-\frac{e}{2}\sum_{j}\left\{  \bm{v}_j, \delta(\bm{r}-\bm{r}_j)  \right\}, 
\\
\bm{j}_{E}^{(0)}(\bm{r})  & =-\frac{1}{2e}\left\{  H , \bm{j}^{(0)}(\bm{r})  \right\}, 
\end{align}
where  $\bm{r}_j$ denotes the position of the $j$th electron, $-e$ ($e>0$) is the electron charge, $\bm{v}_j$ is the velocity operator of the $j$th electron, $H$ is the unperturbed Hamiltonian of the system, and $\{\hat{A},\hat{B}\}=\hat{A}\hat{B}+\hat{B}\hat{A}$ denotes the anticomutator. We note that $\bm{r}_j$ is a quantum mechanical operator, while $\bm{r}$ is a c-number. 

Under the electric potential $\phi(\bm{r})$  and the fictitious gravitational potential $\psi (\bm{r})$, the Hamiltonian is written as $H_{\text{tot}}=H+e\sum_j \bm{r}_j \cdot \nabla \phi  +\frac{1}{2}\left\{  H , \frac{1}{c^2} \sum_j \bm{r}_j \cdot \nabla  \psi (\bm{r}) \right\} $, where $H$ is an unperturbed Hamiltonian and $c$ is the speed of light. Subsequently, the electric current operator $\bm{j}(\bm{r})$ and the energy current operator $\bm{j}_{E}(\bm{r})$ deviate from the equilibrium state. They are written as: 
\begin{align}
	\bm{j}(\bm{r})	&=\bm{j}^{(0)}(\bm{r})+\bm{j}^{(1)}(\bm{r}) 
	\notag \\
	&=\bm{j}^{(0)}(\bm{r})+\frac{1}{2}\left\{ \bm{j}^{(0)}(\bm{r}) , \frac{1}{c^2}
	\sum_{j}\bm{r}_j\cdot
	\nabla \psi (\bm{r})\right\} ,	\label {eq current2} \\
	\bm{j}_E(\bm{r})	&=\bm{j}_{E}^{(0)}(\bm{r})+\bm{j}_{E}^{(1)}(\bm{r}) \notag \\ 
							&=\bm{j}_{E}^{(0)}(\bm{r})+\frac{1}{2}\left\{ \phi (\bm{r}_j) , \bm{j}^{(0)}(\bm{r}) \right\}
	\notag \\						
	&+\frac{1}{4c^2} \sum_{j}
\left(    \left\{   \left\{ H,\bm{r}_j\cdot\nabla \psi (\bm{r})
 \right\}  ,  \bm{j}^{(0)}(\bm{r})  \right\}                          
\right.
\notag \\
&\left.
 +   \left\{   \left\{ \bm{j}^{(0)}(\bm{r}), \bm{r}_j\cdot\nabla \psi (\bm{r})
\right\}  ,  H   \right\} \right)          \label {eq heat current2}.    
\end{align}

The linear response for the electric current and energy current is written as 
\begin{align}
	\bm{J}&=(L^\text{F})_{11}\left[ \bm{E}+\frac{T}{e}\nabla \left(\frac{\mu}{T} \right) \right]
              + (L^\text{F})_{12}\left[ T \nabla\left(\frac{1}{T}\right) - \frac{\nabla\psi}{c^2} \right]  
	\label {eq current}, 
	\\	
	\bm{J}_E&=(L^\text{F})_{12}\left[ \bm{E}+\frac{T}{e}\nabla \left( \frac{\mu}{T} \right) \right] 
                 + (L^\text{F})_{22}\left[ T \nabla\left(\frac{1}{T}\right) - \frac{\nabla \psi}{c^2} \right]  
	\label {eq energy current},
\end{align}
where ``F" is a label which means a fermion, $\bm{E}$ is an electric field,  $\mu$ is the chemical potential, and $(L^F)_{ij}$ is the transport coefficients ($i,j=1,2$). The measurable current densities $\bm{J}$ and $\bm{J}_E$ are obtained by taking average over the volume of the sample and the quantum-mechanical and thermodynamic averages of the current operators $\bm{j}(\bm{r})$ and $\bm{j}_E(\bm{r})$, respectively. Due to the deviations $\bm{j}^{(1)}(\bm{r})$ and $\bm{j}_{E}^{(1)}(\bm{r})$, the thermal transport coefficients $(L^\text{F})^{\alpha \beta}_{ij}$ ($\alpha, \beta=x, y$) consist of two parts, $(S^\text{F})^{\alpha \beta}_{ij}$ and $(M^\text{F})^{\alpha \beta}_{ij}$: 
\begin{widetext}
\begin{align}
	(S^F)^{\alpha \beta}_{ij}  	&=	\frac{i\hbar}{V} \int f(\eta) \text{Tr} \left( j^{\alpha}_i \frac{dG^+}{d\eta} j^{\beta}_j  \delta (\eta-H) -  j^{\alpha}_i  \delta (\eta-H)  j^{\beta}_j  \frac{dG^-}{d\eta} \right) d\eta , \label{eq Sij} \\
	(M^F)^{\alpha \beta}_{11}  	&=	0 , 
	\ \ \ 
	(M^F)^{\alpha \beta}_{12}  	=	-\frac{e}{2V} \int f(\eta) \text{Tr} [\delta (\eta-H) (r^{\alpha}v^{\beta}-r^{\beta}v^{\alpha})] d\eta 
	\label{eq M12}, \\
	(M^F)^{\alpha \beta}_{22}  	&=	 \frac{1}{V}\int \eta f(\eta) \text{Tr} \delta (\eta-H) (r^{\alpha}v^{\beta}-r^{\beta}v^{\alpha}) d\eta 
										+ \frac{i\hbar}{4V}\int f(\eta) \text{Tr} \delta (\eta-H) [v^{\alpha} , v^{\beta}] d\eta \label{eq M22}. 
\end{align}
\end{widetext}
Here $G^{\pm}$ is the Green's function $G^{\pm}(\eta)=(\eta-H\pm i\epsilon )^{-1}$ which is introduced in Eq.~\eqref{eq Sij} via $\delta(\eta-H)=-(G^+ - G^-)/2\pi i$, $f(\eta)$ is the Fermi distribution function $f(\eta)=\left(e^{\beta\left(\eta-\mu \right)}+1\right)^{-1}$, $\bm{j}_{1}=-e\bm{v}$, $\bm{j}_{2}=\frac{1}{2}(H\bm{v}+\bm{v}H)$, and $\bm{v}$ is the velocity of electrons. 
$(S^F)^{\alpha \beta}_{ij}$ is calculated from the current operators in equilibrium state, $\bm{j}^{(0)}(\bm{r})$ and $\bm{j}_{E}^{(0)}(\bm{r})$, with the deviation of the distribution function from the equilibrium state; $(M^F)^{\alpha \beta}_{ij}$ is calculated from the deviation of the current operators, $\bm{j}^{(1)}(\bm{r})$ and $\bm{j}_{E}^{(1)}(\bm{r})$, with the equilibrium distribution function.  
Actually, $(S^F)^{\alpha \beta}_{ij}$ is the Kubo formula, and $(M^F)^{\alpha \beta}_{ij}$ represent correction terms. The total thermal transport coefficients are their sums: $(L^F)^{\alpha \beta}_{ij}=(S^F)^{\alpha \beta}_{ij}+(M^F)^{\alpha \beta}_{ij}$. 

From these results, we can derive some useful equations for later calculations. First, we can write down $(S^F)^{\alpha \beta}_{ij}$ in terms of the Berry phase. For example, $(S^F)^{\alpha \beta}_{12}$ is written as: 
\begin{align}
	(S^F)^{\alpha\beta} _{12} =-\frac{e}{\hbar V} \text{Im}\sum_{n,\bm{k}} 
f( \varepsilon_{n\bm{k}}) \left\langle \frac{\partial u_{n}}{\partial k_{\alpha}} \right|   \left( H+\varepsilon _{n\bm{k}} \right)    \left|  \frac{\partial u_{n}}{\partial k_{\beta}} \right\rangle. 
	\label{eq SA12}
\end{align}   
Second, because the Fermi distribution function $f(\eta)$ becomes the step function $\Theta(\mu-\eta)$ in the zero temperature limit,  $(L^F)^{\alpha \beta}_{12}$ and $(S^F)^{\alpha \beta}_{12}$ is written as 
\begin{align}
(L^F)^{\alpha \beta}_{12} 	&=	 \frac{\mu}{-e} (L^F)^{\alpha\beta}_{11} 
\notag \\
&= - \frac{2e\mu}{\hbar V} \text{Im} \sum_{n,\bm{k}}  \Theta(\mu-\varepsilon_ {n\bm{k}}  )\left\langle \frac{\partial u_n }{\partial k_{\alpha}}   \left|  \frac{\partial u_n }{\partial k_{\beta}} \right. \right\rangle ,
\end{align} 
and
\begin{align}
	(S^F)^{\alpha\beta} _{12} 	
&= 	-\frac{2e}{\hbar V} \text{Im} \sum_{n,\bm{k}}    \Theta(\mu-\varepsilon _{n\bm{k}}   )  
\notag \\
&\times \left\langle \frac{\partial u_n }{\partial k_{\alpha}} \left|   \left( \frac{H+\varepsilon _{n\bm{k}} }{2}\right)    \right|  \frac{\partial u_n }{\partial k_{\beta}} \right\rangle.
\end{align} 
Thus the relation $(L^F)^{\alpha \beta}_{12}=(S^F)^{\alpha \beta}_{12}+(M^F)^{\alpha \beta}_{12}$ and Eq.~\eqref{eq M12} in the zero temperature limit lead to the following useful formula: 
\begin{align}
	&\text{Tr} [\delta (\mu-H) (r^{\alpha}v^{\beta}-r^{\beta}v^{\alpha}) ]	
	\notag \\
	&= \frac{d}{d\mu}\int_{-\infty}^{\mu} \text{Tr} [\delta (\eta-H) (r^{\alpha}v^{\beta}-r^{\beta}v^{\alpha}) ]d\eta \notag 
	\\	
	&=\frac{2V}{-e} \frac{d}{d\mu} \left. (M^F)^{\alpha\beta} _{12} \right |_{T\rightarrow 0} \notag \\
	&=-\frac{2}{\hbar } \frac{d}{d\mu} \sum_{n,\bm{k}} \Theta (\mu-\varepsilon _{n\bm{k}} ) 
	\notag \\ 
	&\times \text{Im} \left\langle \frac{\partial u_n }{\partial k_{\alpha}} \right|   \left(H+\varepsilon _{n\bm{k}} - 2\mu \right)    \left|  \frac{\partial u_n }{\partial k_{\beta}} \right\rangle . \label{eq Trrv}
\end{align}
We note that this equation does not depend whether the particles are fermion or boson. Therefore we can apply this equation to the magnon system as well.

\end{document}